\documentclass[times, twocolumn]{aastex631}
\usepackage{CJK}
\usepackage{graphicx}
\usepackage{enumerate}
\usepackage{amssymb, amsmath}
\usepackage{color}
\usepackage{ulem}
\usepackage{dblfnote}
\usepackage{appendix}
\usepackage[stable]{footmisc}
\usepackage{comment}
\usepackage{footnote}
\usepackage[nodayofweek]{datetime}
\newdateformat{monthyearday}{%
  \THEYEAR\ \monthname[\THEMONTH] \twodigit{\THEDAY}}

\newcommand{\Teff}{\ensuremath{T_{\rm eff}}}

\newcommand{\mj}{\ensuremath{\,M_{\rm J}}}

\newcommand{\mum}{$\mu$m}
\newcommand{\wbd}{WISE 1828+2650}

\usepackage{longtable}

\begin{document}
\shorttitle{WISE 1828+2650}
\shortauthors{De Furio et al.}

\accepted{February 21, 2023}
\submitjournal{ApJ}


\title{JWST Observations of the Enigmatic Y Dwarf \wbd: I. Limits to a Binary Companion  }

\correspondingauthor{Charles Beichman, chas@ipac.caltech.edu; Matthew De Furio, defurio@umich.edu}


\author[0000-0003-1863-4960]{Matthew De Furio}
\affiliation{Department of Astronomy, University of Michigan, Ann Arbor, MI 48109, USA}
\author[0000-0003-1487-6452]{Ben Lew}
\affiliation{Bay Area Environmental Research Institute, Moffett Field, CA 94035, USA}
\affiliation{NASA Ames Research Center, Mountain View, CA}
\author[0000-0002-5627-5471]{ Charles Beichman}
\affiliation{NASA Exoplanet Science Institute, Infrared Processing and Analysis Center (IPAC)}
\affiliation{Jet Propulsion Laboratory, California Institute of Technology, Pasadena, CA 91125}

\author[0000-0002-6730-5410]{Thomas Roellig}
\affiliation{NASA Ames Research Center, Mountain View, CA}


\author[0000-0001-5966-837X]{Geoffrey Bryden}
\affiliation{Jet Propulsion Laboratory, California Institute of Technology, Pasadena, CA 91125}

\author[0000-0002-5741-3047]{David Ciardi}
\affiliation{NASA Exoplanet Science Institute, Infrared Processing and Analysis Center (IPAC)}
\affiliation{Jet Propulsion Laboratory, California Institute of Technology, Pasadena, CA 91125}

\author[0000-0003-1227-3084]{Michael Meyer}
\affiliation{Department of Astronomy, University of Michigan, Ann Arbor, MI 48109, USA}

\author[0000-0002-7893-6170]{Marcia Rieke}
\affiliation{Steward Observatory, University of Arizona, Tucson, AZ 85721, USA}

\author[0000-0002-7162-8036]{Alexandra Greenbaum}
\affiliation{IPAC, Caltech, 1200 E. California Blvd., Pasadena, CA 91125, USA}
\author[0000-0002-0834-6140]{Jarron Leisenring}
\affiliation{Steward Observatory, University of Arizona, Tucson, AZ 85721, USA}
\author[0000-0002-3414-784X]{Jorge Llop-Sayson}
\affiliation{California Institute of Technology, 1200 E. California Blvd., Pasadena, CA 91125, USA}
\author[0000-0001-7591-2731]{Marie Ygouf}
\affiliation{Jet Propulsion Laboratory, California Institute of Technology, Pasadena, CA 91125}

\author[0000-0003-0475-9375]{Loic Albert}
\affiliation{Universit\'e de Montr\'eal, Montr\'eal,Quebec H3C 3J7, Canada,}
\author{Martha Boyer}
\affiliation{Space Telescope Science Institute, 3700 San Martin Drive, Baltimore, MD 21218, USA}
\author{Daniel Eisenstein}
\affiliation{Harvard-Smithsonian Center for Astrophysics, 60 Garden St, Cambridge, MA 02138, USA}
\author{Klaus Hodapp}
\affiliation{University of Hawaii, Hilo, HI,96720}
\author{Scott Horner}
\affiliation{NASA Ames Research Center, Mountain View, CA}
\author[0000-0002-6773-459X]{Doug Johnstone}
\affiliation{NRC Herzberg Astronomy and Astrophysics, 5071 West Saanich Rd, Victoria, BC, V9E 2E7, Canada}
\affiliation{Department of Physics and Astronomy, University of Victoria, Victoria, BC, V8P 5C2, Canada}
\author{Doug Kelly}
\affiliation{Steward Observatory, University of Arizona, Tucson, AZ 85721, USA}
\author{Karl Misselt}
\affiliation{Steward Observatory, University of Arizona, Tucson, AZ 85721, USA}
\author{George Rieke}
\affiliation{Steward Observatory, University of Arizona, Tucson, AZ 85721, USA}
\author[0000-0003-2434-5225]{John Stansberry}
\affiliation{Space Telescope Science Institute, 3700 San Martin Drive, Baltimore, MD 21218, USA}
\author[0000-0002-6395-4296]{Erick Young}
\affiliation{Universities Space Research Association, 425 3rd Street SW, Suite 950, Washington DC 20024}




\begin{abstract}
The Y-dwarf WISE 1828+2650 is one of the coldest known Brown Dwarfs with an effective temperature of $\sim$300 K. Located  at  a distance of just 10 pc, previous model-based estimates suggest WISE1828+2650 has a mass of $\sim$5-10 \mj, making it a valuable laboratory for understanding the formation, evolution and physical  characteristics of gas giant planets. However, previous photometry and spectroscopy have presented a puzzle with the near-impossibility of simultaneously fitting both the short (0.9-2.0 \mum) and long wavelength (3-5 \mum) data. A potential solution to this problem has been the suggestion that \wbd\ is a binary system whose composite spectrum might provide a better match to the data. Alternatively, new models  being developed to fit JWST/NIRSpec and MIRI spectroscopy might provide new insights. This article describes JWST/NIRCam observations of \wbd\ in 6 filters to address the binarity question and to provide new photometry to be used in model fitting.  We also report Adaptive Optics imaging with the Keck 10 m telescope. \textit{We find no evidence for multiplicity for a companion beyond 0.5 AU with either JWST or Keck}. Companion articles will present low and high resolution spectra of WISE 1828 obtained with  both NIRSpec and MIRI. 
\end{abstract}


\section{Introduction}
 
 The WISE mission \citep{Wright2010} identified cool brown dwarfs (BDs) using their very red ([3.4]--[4.6] or hereafter W1-W2) colors \citep{Kirkpatrick2011}. The most extreme of these objects, with  effective temperatures  \Teff\  $<$500 K,  have been typed as Y dwarfs \citep{Cushing2011,Kirkpatrick2012}.  With  W1-W2$>$4.2 mag, WISEP J182831.08+265037.8 (hereafter \wbd) was identified as one of the reddest  of this small group  of only two dozen Y dwarfs. Follow-up photometry  and spectroscopy  with the Hubble Space Telescope (HST), the Spitzer Space Telescope, and various ground-based facilities found weak emission between 1.0-1.7 \mum\ marked by absorption due to H$_2$O, CH$_4$ and NH$_3$ \citep{Cushing2011}, extremely weak emission in the $K$ band (2.2 \mum), followed by a sharp rise out to 5 \mum.
 
 Astrometry from Keck, Spitzer and HST determined distances to the WISE Y dwarf sample ranging from 5-15 pc \citep{Beichman2013,Martin2018,Kirkpatrick2019}. With  distances serving to constrain the absolute luminosity and  typical ages of 3-10 Gyr inferred from  their transverse  motions, it was possible to fit  the  photometry and spectroscopy to evolutionary models to estimate masses from  5-10 \mj.   \citet{Beichman2014} and \citet{Kirkpatrick2019} put \wbd\  at a distance of of 9.93$\pm$0.23 pc and suggested an effective temperature of 400K and a mass around 5 \mj.

\begin{deluxetable*}{lllllll}
\tablewidth{0pt}
\tablecaption{JWST NIRCam Observing Parameters (PID:\#1189)\label{tab:exposures}
}
\tablehead{
\colhead{Instrument} & \colhead{Filter Pair}
&\colhead{Readout} & \colhead{Groups/Int} & \colhead{Ints/Exp}& \colhead{Dithers} & \colhead{Total Time (sec)}}
\startdata
NIRCam &F115W \& F335M (CH$_4$)& MEDIUM&4&10&4 &1632\\
NIRCam &F140M (H$_2$O, CH$_4$) \& F360M&MEDIUM&4&10&4 &1632\\
NIRCam &F162M (ref for F140M) \& F470N (H$_2$) &MEDIUM&4&10&4 &1632\\ 
\enddata
\end{deluxetable*}

 \wbd\ has proven to be exceptional even within the  unusual class of Y dwarfs. Its extreme W1-W2 color, the difficulty in simultaneously fitting models to the 1-2 and 3-5 \mum\ photometry,  and \wbd's position at least 1  mag above the Y/T dwarf locus in the H-W2 color-magnitude diagram \citep[Figure 8]{Kirkpatrick2019} highlighted the challenges of making satisfactory models for such cold objects. \citet{Leggett2013,Leggett2017} suggested that \wbd\  might be a binary system which would account for at least 0.75 mag of its separation from the Y/T locus. Most recently, \citet{Cushing2021} used 0.7-1.7 \mum\ HST spectroscopy and a  new generation of models \citep[the Sonora Bobcat models]{Marley2021} to bolster the idea that \wbd\  is an unresolved binary consisting of roughly equal mass objects with \Teff$\sim$275-350 K. However, there remain challenges with the model ages  ($<$1 Gyr for a single object)  and the subsolar values of [M/H] $\sim-0.5$ and [C/O] $\sim-0.6$ required to fit the existing data \citep{Cushing2021}.
 
With the expectation that JWST would help resolve these theoretical difficulties, we set out a program of imaging and spectroscopy, particularly in the key 3-10 \mum\ region. The JWST NIRCam imaging data  permit a search for faint companions at  separations and sensitivity levels not previously possible as well as anchoring the fluxes of the the higher-spectral resolution NIRSpec data. The spectroscopy provides critical diagnostics  of physical conditions, composition, the presence or absence of clouds, and  surface gravity in the 3-5 \mum\ region  where Y dwarfs emit most of their energy.  
 
 This paper describes NIRCam  imaging in six filters spanning 1-5 \mum\ undertaken by the NIRCam Guaranteed Time Observation (GTO) team under PID\#1189. Companion papers will give results from  NIRSpec and MIRI spectroscopy, and from a deep  NIRISS search for a close companion. $\S$\ref{sec:obs} describes the observations and $\S$\ref{sec:data} the data reduction procedures and results. $\S$\ref{sec:companions}  addresses the search for and limits to the presence of a close companion to \wbd\ while $\S$\ref{sec:SED} presents fits to the spectral energy distribution of \wbd\ and estimates of derived physical parameters. The paper ends  with a search for wide field companions ($\S$\ref{sec:widefield}) and concluding remarks. 
 
\begin{deluxetable}{lll}[t!]
\tablewidth{0pt}
\tablecaption{NIRCam Photometry \label{tab:data}
}
\tablehead{
\colhead{Filter}&   \colhead{F$_\nu(\mu$Jy)$^1$}&\colhead{Magnitude}}
\startdata
F090W&0.078$\pm$0.005(0.015)&26.16$\pm$0.20\\
F115W&0.311$\pm$0.005(0.03)&24.39$\pm$0.10\\
F162M&1.29$\pm$0.009(0.13)&22.25$\pm$0.10\\
F335M&2.41$\pm$0.013(0.24)&20.23$\pm$0.10\\
F360M&34.2$\pm$0.05(3.4)&17.19$\pm$0.10\\
F470M&340.4$\pm$0.62(34)&14.16$\pm$0.10\\
F480M$^2$&326$\pm$0.14(32)&14.21$\pm$0.10\\
\hline
J (1.25 \mum)&0.65$\pm$0.23&23.48$\pm$0.23\\
H (1.65 \mum)&0.83$\pm$0.21&22.73$\pm$0.13\\
K (2.16 \mum)&0.27$\pm$0.08&23.48$\pm$0.36\\
Spitzer/IRAC (3.55 \mum)&48.1$\pm$1.0&16.92$\pm$0.02\\
Spitzer/IRAC (4.55 \mum)&335.5$\pm$6.71&14.32$\pm$0.02\\
WISE W2 (4.62 \mum)&310$\pm$14&14.35$\pm$0.05\\
\enddata
\tablecomments{$^1$ First quoted uncertainty reflects only photometric accuracy based on aperture photometry. Values in parenthesis include an average 10\% calibration uncertainty at all wavelengths  estimated from the average of  successive iterations of the calibration values  (PHOTMJYSR and PIXAR\_SR)  available at MAST. The values listed here use image files created  on 2022-11-11 using  calibration software version 11.16.14 and photometric reference data from jwst\_nircam\_photom\_0114.fits.  Values below the solid line come from ground and space missions as given in \citep{Martin2018,Liu2016,Kirkpatrick2019,Kirkpatrick2021}. $^2$Observation obtained using the NIRISS instrument}.
\end{deluxetable}

\begin{figure*}%
\includegraphics[width=0.45\textwidth]{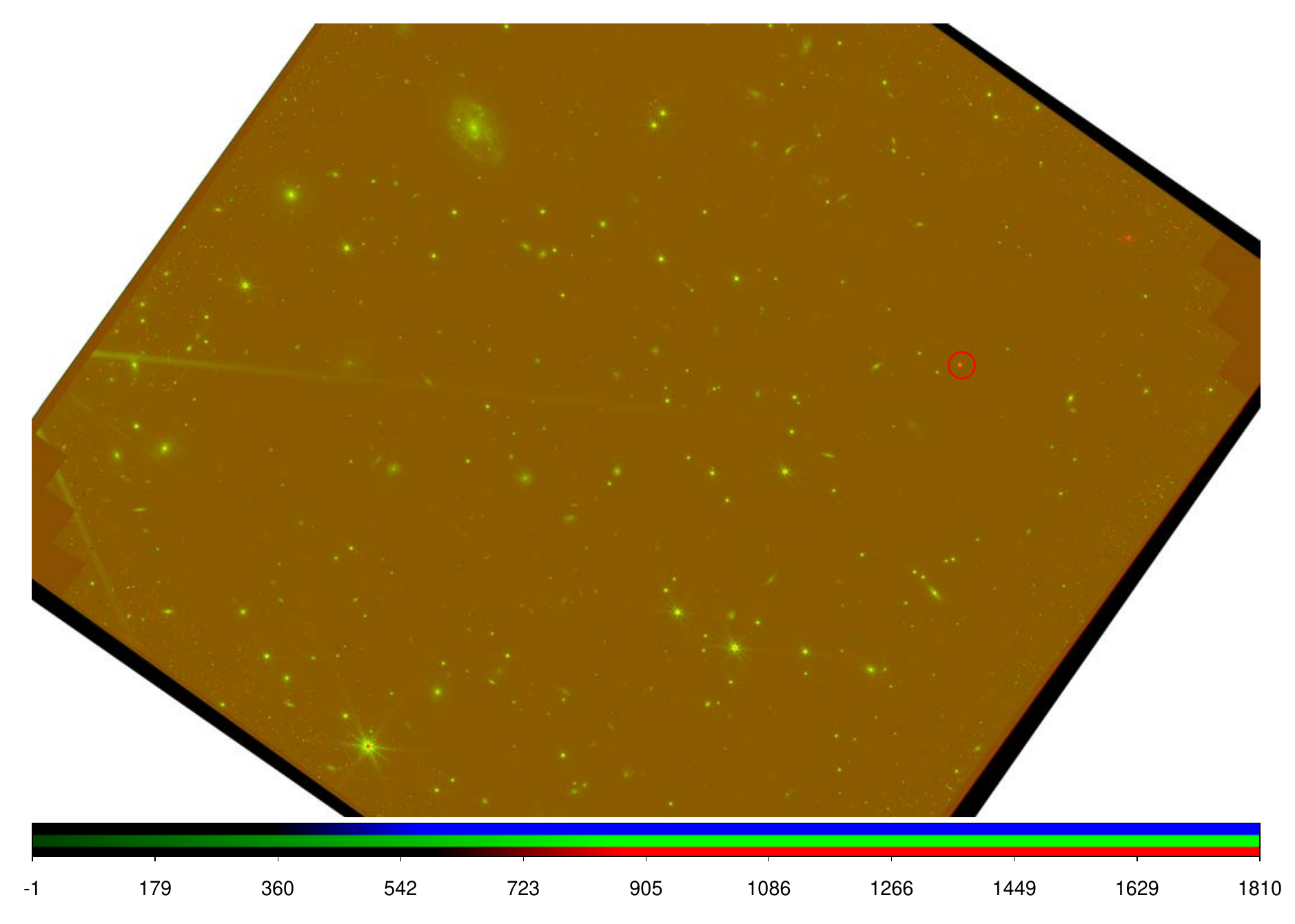}\includegraphics[width=0.45\textwidth]{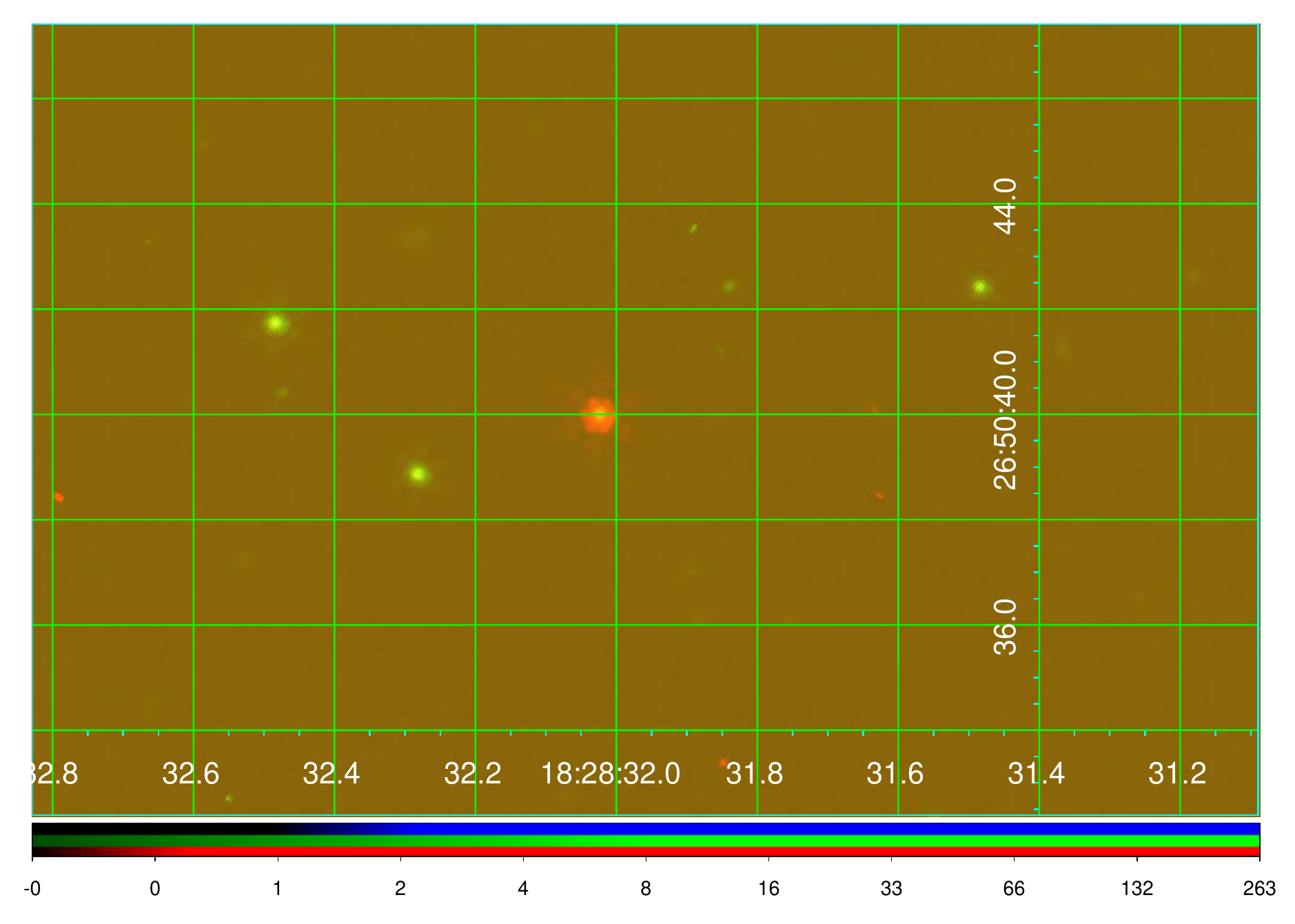}  
\caption{left) A color-composite image of 3 NIRCam filters, F162M, F335M and F470N showing one extremely red object at the expected position of \wbd. right) a zoom-in on the position of \wbd.  \label{fig:image}}
\end{figure*}

\section{Observations and Data Reduction \label{sec:obs}}

Table~\ref{tab:exposures} describes the observing parameters for this program, using the long and short wavelength modules of NIRCam to observe  \wbd\  simultaneously in  a combination of narrow, medium and wide filters between 0.9 and 4.7 \mum. Data were obtained on 2022-07-28 UTC. The NIRISS full frame imager was used to obtain a long exposure in the F480M filter in a deep search for a close companion. The NIRISS results will be presented separately, but the photometry at that wavelength is presented here for comparison with models.  NIRSpec and MIRI spectroscopy were also obtained at this time and are discussed separately. 

NIRCam images in the six wavelength bands and the single NIRISS band were downloaded from the Mikulski Archive for Space Telescopes (MAST) at the Space Telescope Science Institute, specifically the "...i2d.fits" files which are the result of mosaicing the four dither positions obtained at each wavelength. The specific observations analyzed can be accessed via \dataset[DOI]{https://archive.stsci.edu/doi/resolve/resolve.html?doi=10.17909/jxva-9x26}. Figure~\ref{fig:image} shows a three- color, full-frame NIRCam image of the field around \wbd\ and is composed of images obtained at F162M (blue),  F335M (green) and F470M (red) with a close-up around the position of \wbd. Only one object in the frame appears with the highly red color characteristic of a cool Y dwarf.


\section{Analysis \label{sec:data}}

The values listed here use image files created  on 2022-11-11 using  calibration software version 11.16.14 and photometric reference data from jwst\_nircam\_photom\_0114.fits.

\subsection{Photometry}
Table~\ref{tab:data} shows the results of standard aperture photometry using the astropy/photutils package \citep{Astropy2013, Astropy2018, Astropy2022,Bradley2020} with the image files created on 2022-11-11 and the calibration software version 11.16.14 with photometric reference data from jwst\_nircam\_photom\_0114.fits. We selected  an aperture size corresponding to  70\% encircled energy and the appropriate aperture correction from the CRDS database\footnote{https://jwst-crds.stsci.edu/}. The F090W filter was the weakest detection with the total flux dependent on the selected aperture due to the poorly defined Point Spread Function.  For small apertures with encircled energy $<$ 50\%, the flux estimates in F090W were variable at the 20-30\% level. Beyond encircled energy of 50\%, the flux estimates were consistent at the 10\% level, shown in the total uncertainty in Table \ref{tab:data}. We adopted a large radius of 9.25 pixels encircling 85\% of the total energy for the F090W filter. We also obtained photometry using the NIRISS imager in full frame mode \citep{Willott2022} using the same procedure as described above.

\begin{figure*}[t!]%
\centering
\includegraphics[width=0.9\textwidth]{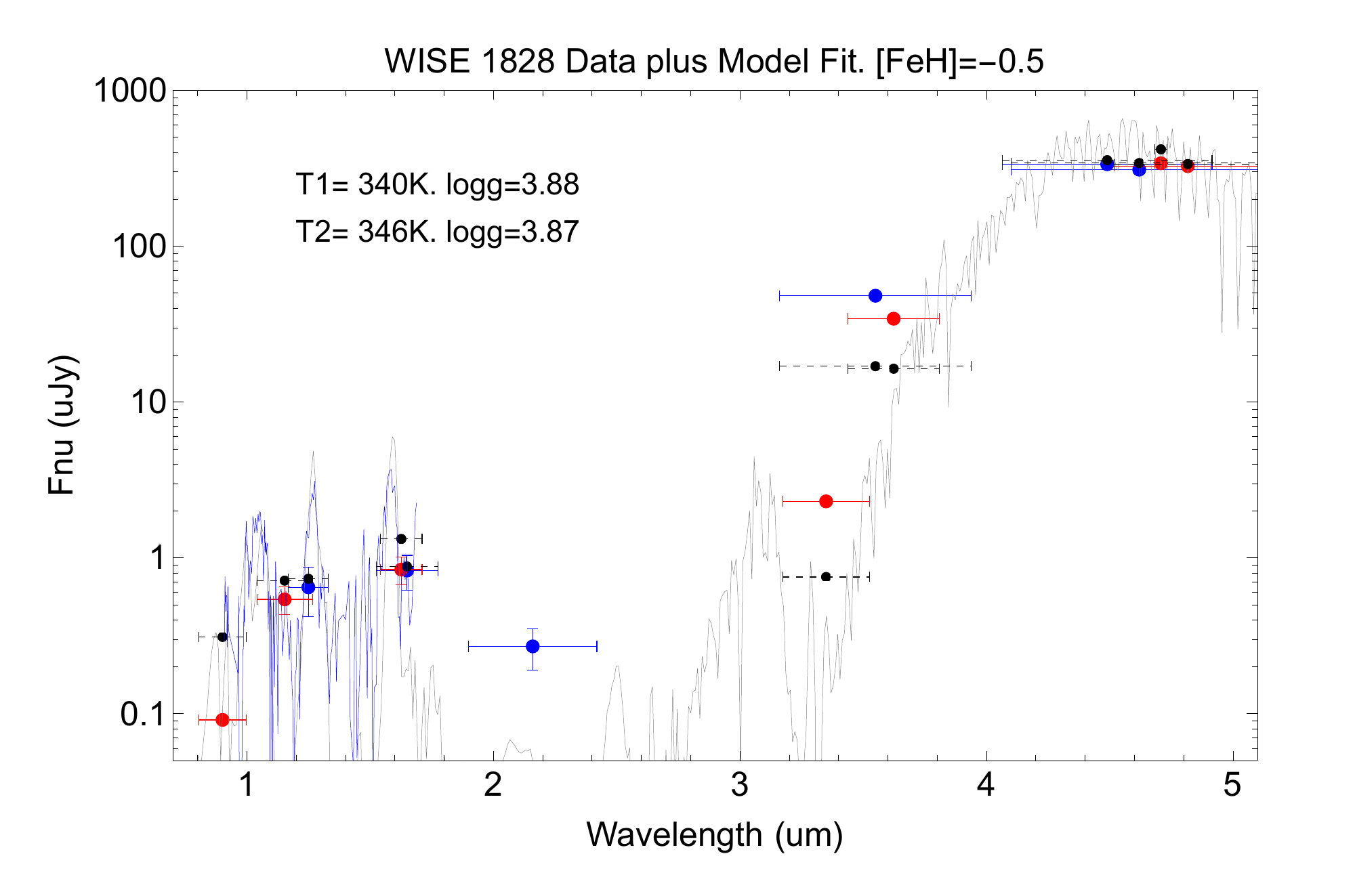} 
\caption{The figure shows a combination of the new JWST photometry (large red circles), previous ground- based and space-based observations (large blue circles)   \citep{Kirkpatrick2011,Leggett2015,Kirkpatrick2019}, and in  the  1-2 \mum\ region  the HST spectrum \citep[thin blue line]{Cushing2021}. Horizontal bars denote the widths of the various filters.  As discussed further in $\S$~\ref{sec:SED}, we also  plot as a thin grey line a cloudless Sonora (Bobcat)  spectrum  smoothed to R$\sim$3000  for the best fitting,  binary object solution (Table~\ref{tab:Sonora}).  The black circles with dashed lines denote predicted fluxes  as integrated over the relevant passbands for each filter \citep{Marley2021}.  \label{fig:photometry}}
\end{figure*}

Figure \ref{fig:photometry} shows the JWST results along with previous data from Spitzer, WISE, HST and Keck \citep{Kirkpatrick2011, Leggett2015, Kirkpatrick2019, Cushing2021}.  \wbd\ is  detected at high signal-to-noise (SNR) in all  filters with statistical uncertainties of just a few percent. However, we adopt a higher level of uncertainty for the absolute calibration in the model fitting. Successive MAST releases of these data from August-November 2022 have shown significant variations in the absolute calibration of the NIRCam filters used here. We adopt  a uniform  10\% uncertainty for the absolute photometric calibration in all filters. 

We note the consistency between the JWST data and previous observations although different passbands can be expected to yield significant differences given the highly structured nature of the SED of cool BDs. The close agreement between multiple facilities at the wavelengths of peak emission around 4-5 \mum\ is consistent with the general lack of variability in the brightness of \wbd.  Brooks et al.  (in preparation) investigated the variability of  several hundred cold BDs observed  in the    \citet{Kirkpatrick2021} astrometric program using Spitzer. For \wbd\ they found a  limit of 3\% ($3\sigma$) to the variability in IRAC Ch2 ([4.5]) in  28 observations spanning almost 8 years.  A model fit based on the cloudless Sonora-Bobcat  models \citep{Marley2021} is  also plotted and  will be discussed in $\S$~\ref{sec:SED}.

\subsection{Astrometry \label{sec:astrometry}}
As described in \citep{Kirkpatrick2019},\wbd\ has a high proper motion ($\sim$1\arcsec/yr), mostly in right ascension,  and  a  parallax corresponding to a distance of 9.93$\pm$0.23 pc. The values listed in Table~\ref{tab:posn} were used  to establish the pointing of the NIRCam, NIRSpec and MIRI observations in this program.

The F360M image was used to obtain the position of \wbd. The high signal to noise ratio in this  filter would imply a nominal positional accuracy of FWHM/(2$\times$SNR) $<$ 5 milli-arcseconds (mas) where FWHM is the Full Width at Half Maximum of the Point Spread function. We calculated both center-of-mass and  2-D Gaussian fits to obtain the image centroid which yielded nominal  accuracy  of $<$0.1 pixel or $<$6 mas. Information in the FITS header was used to convert the pixel location into celestial coordinates (J2000, 2022.5699) as given in Table~\ref{tab:posn}\footnote{The astrometric  distortion correction in the MAST processing used the file crds://jwst\_nircam\_distortion\_0158.asdf}

\begin{figure}[t!]%
\includegraphics[width=0.45\textwidth]{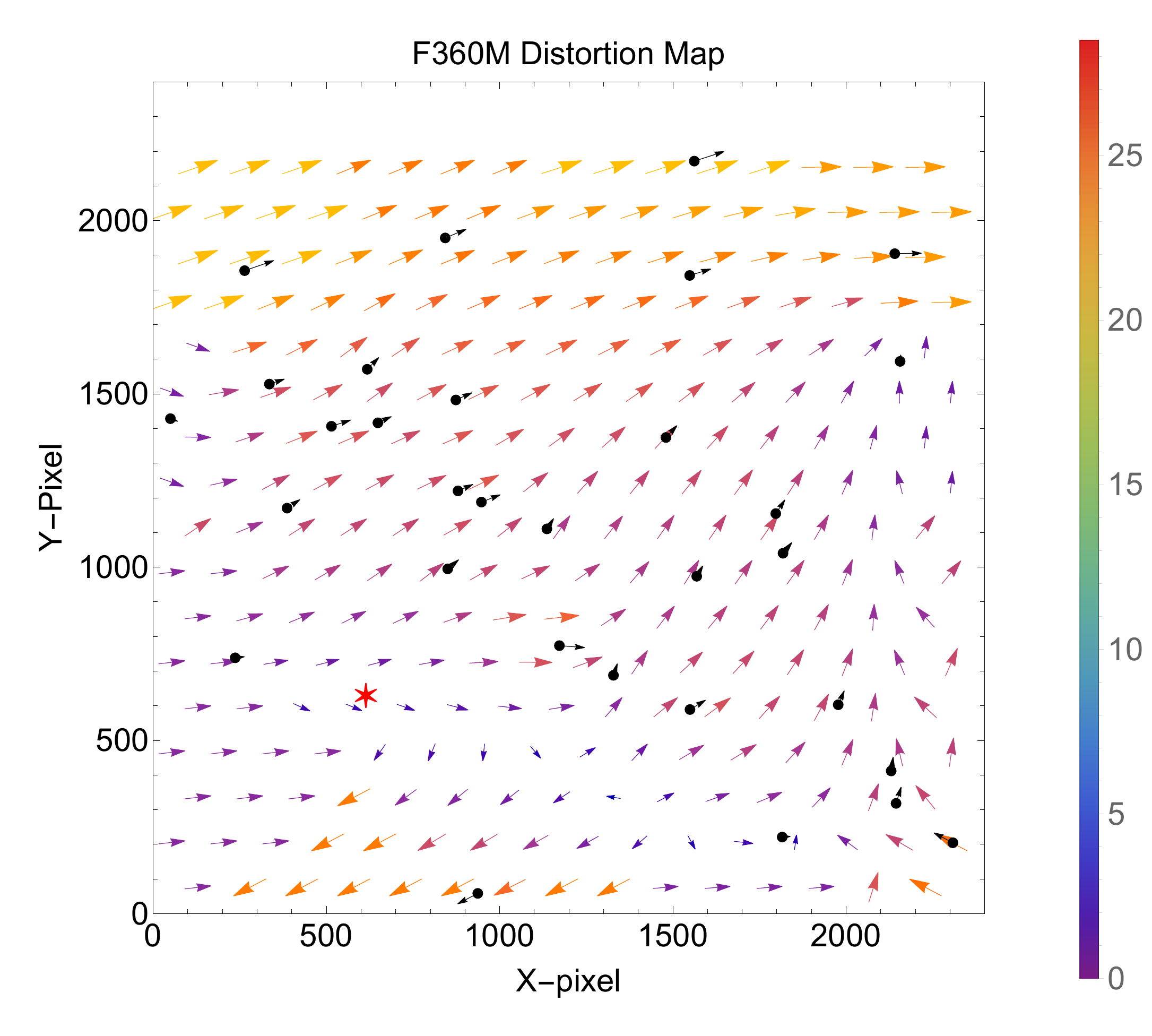} 
\caption{Differences in the positions of 31 Gaia DR3 stars (selected for negligible parallactic motions and  corrected for proper motion) relative to the positions observed by JWST in F360M.  Individual Gaia stars are shown as black symbols with the arrow showing the difference in pixels (63 mas), scaled by a factor of 200 for visibility. The position of \wbd\ is shown as a red star. The colors in the image and the color bar encode the magnitude of deviation in mas. The maximum Gaia-JWST difference is 27 mas with an average of 10 mas and dispersion of 10 mas.\label{fig:distort}}
\end{figure}

A comparison of the NIRCam positions of 31    unsaturated, proper-motion corrected Gaia stars ($20\, {\rm mag} >{\rm Gmag}>$14  mag) across the 4\arcmin$\times$4\arcmin\ field  showed  a small offset of order the measurement uncertainty with dispersion of $(\Delta\alpha,\Delta\delta)$= 10$\pm$10, 7$\pm$6) mas, demonstrating that the reference frame is accurate to $\sim$5-10 mas. However,  Figure~\ref{fig:distort} shows that there remains some coherent distortions  across the field of view.

The predicted  position of \wbd\  was calculated at the JWST epoch incorporating  stellar  parallax \citep{Smart1977} using the rectangular coordinates of JWST's location in the solar system as given in the FITS header (reversed to give the coordinates of the Sun).   The uncertainty in the F360M position is the combination of the centroid uncertainty and the uncertainty in the reference frame  derived from the Gaia stars, $\pm$2 mas, for a combined uncertainty of $\pm$4.4 mas. The difference in the predicted vs observed positions  is within the  uncertainties derived from a Monte Carlo run in which the \citet{Kirkpatrick2021} values were varied according to their quoted uncertainties. The Monte Carlo distribution  yielded   uncertainties at the JWST epoch  of 6.4 mas in right ascension and  declination. The Predicted-JWST position difference is   consistent  with the existing parallax and proper motion values for \wbd.

\begin{deluxetable*}{lllllll}
\tablewidth{0pt}
\tablecaption{NIRCam Astrometry of WISE 1828 \label{tab:posn}
}
\tablehead{\colhead{Observatory}&
\colhead{Epoch}& \colhead{RA } & \colhead{Dec}
&\colhead{Parallax}&\colhead{$\mu_{RA}$}&\colhead{$\mu_{DEC}$}\\
&\colhead{(MJD)}& \colhead{RA (Equinox=J2000)} & \colhead{(Equinox=J2000)}
&\colhead{(mas)}&\colhead{(mas/yr)}&\colhead{(mas/yr)}}
\startdata
Spitzer$^1$&57094.09&277.131096 ($\pm$2 mas)&26.844069($\pm$2 mas)&100.3$\pm$2&1016.5$\pm$0.8&169.3$\pm$0.8\\
&&18$^h$28$^m$31.463$^s$&+26$^o$50$^\prime$38.65\arcsec&&&\\
JWST (predicted)&59788.5143&277.1334156 ($\pm$6 mas)&26.8444348 ($\pm$6 mas)&&&\\
&&18$^h$28$^m$32.020$^s$&+26$^o$50$^\prime$39.965\arcsec&&&\\
JWST (F360M)&59788.5143&277.1334164 ($\pm$10 mas)&26.8444334 ($\pm$10 mas)&&\\
&&18$^h$28$^m$32.020$^s$&+26$^o$50$^\prime$39.960\arcsec&&&\\ \hline
$\Delta$Spitzer-JWST(pred) &&$-$7.45\arcsec&$-$1.32\arcsec&&&\\
$\Delta$JWST (Pred-Obs)&& 2.5$\pm$12 mas & -5.1$\pm$8.2 mas&&&\\ \hline
Gaia-JWST differences$^3$ (mas)&&10.4$\pm$10.2 (1.8) &6.8$\pm$5.1 (0.9) \\
\enddata
\tablecomments{$^1$\citet{Kirkpatrick2019}. $^2$Absolute astrometric accuracy estimated by measurements of  nearby Gaia stars.
$^3$ Difference in positions between 31 proper-motion corrected Gaia stars and their JWST values. The value in parenthesis is $\sigma\_{mean}$ for the sample and represents an estimate of the overall precision of reference frame.}
\end{deluxetable*}


\section{Search for Close Companions \label{sec:companions}}

\subsection{Near-IR Observations}
Previous observations with Keck and HST revealed no evidence for multiplicity at the level of 0.1\arcsec\ \citep{Beichman2013}, corresponding to orbital separations of $>$1 AU for equal brightness components. However, the Keck observations were done with the 40 mas/pixel wide-field camera mode with a typical resolution of $\sim0.16\arcsec.$  New Keck observations were obtained 2022-Sep-08 UT with NIRC2 behind the laser guide star adaptive optics system \citep{Wizinowich2000} in the narrow-field camera mode with a pixel scale of $0.009942\arcsec$.  

Observations of WISE 1828+2650 were made in the $H$ filter $(\lambda_o = 1.633; \Delta\lambda = 0.296~\mu$m) with an integration time of 300 seconds per frame in a standard 3-point dither pattern that is used with NIRC2 to avoid the noisier lower-left quadrant.  The transparency and seeing during the night was highly variable and not all frames detected the BD. A total of 12 frames at 300 seconds each were acquired yielding a total on-source integration time of 3600 seconds.

The science frames were flat-fielded and sky-subtracted.  The flat fields were generated from a median average of dark subtracted flats taken on-sky.  The flats were normalized such that the median value of the flats is unity.  The sky frames were generated from the median average of the dithered science frames; each science image was then sky-subtracted and flat-fielded.  The reduced science frames were combined into a single combined image using a intra-pixel interpolation that conserves flux, shifts the individual dithered frames by the appropriate fractional pixels, and median-coadds the frames.  The final resolutions of the combined dithers was determined from the full-width half-maximum of the point spread functions: 7.1 pixels = 0.071\arcsec.

\begin{figure}
    \centering
    \includegraphics[width=0.5\textwidth]{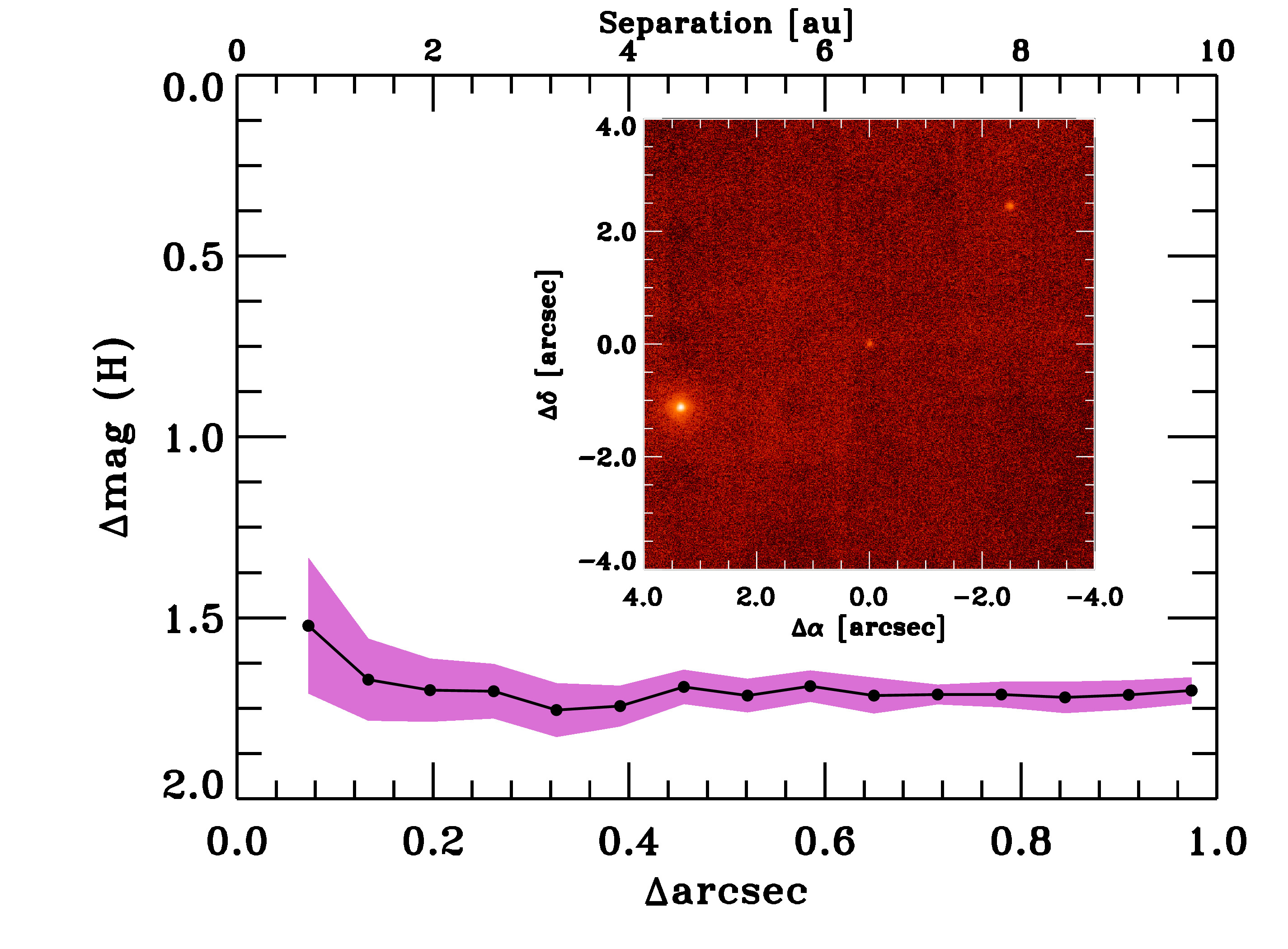}
    \caption{Keck NIR AO imaging and sensitivity curve for \wbd\ taken in the $H$ filter. We can recover a  companion at a contrast of $\sim 1.5$ magnitudes relative to the BD at separations greater than 0.071\arcsec. {\it Inset:} Image of the central portion of the NIRC2 image.
    }
    \label{fig:ao_image}
\end{figure}

The sensitivity of the final combined AO image was  determined by injecting simulated sources azimuthally around the primary target every $20^\circ $ at separations of integer multiples of the central source's FWHM. The brightness of each injected source was scaled until standard aperture photometry detected it with $5\sigma $ significance. The resulting brightness of the injected sources relative to \wbd\ set the contrast limits at that injection location. The final $5\sigma $ limit at each separation was determined from the average of all of the determined limits at that separation, and the uncertainty on the limit was set by the rms dispersion of the azimuthal slices at a given radial distance.  The Keck data have a sensitivity close-in of $\Delta H \approx 1.5$ mag at 0.071\arcsec; the final sensitivity curve for the Keck image is shown in (Figure~\ref{fig:ao_image}).  No close-in stellar companions were detected and the FWHM of the \wbd\ is consistent with the brighter nearby star to the southeast (Gaia DR3 4585337218702066560). \wbd\ is $\Delta H = 3.4 \pm 0.2$ mag fainter than Gaia DR3 4585337218702066560. 

The $H$-band magnitude of \wbd\ as measured by the Keck data is $H=22.9\pm0.2$ mag. Using the JWST F162M filter as a proxy for the H-band filter, this measurement is consistent with the infrared photometry determined directly from the JWST images (see Table~\ref{tab:data}).


\subsection{NIRCam Observations\label{NIRCamObs}}

Although the aperture of JWST is smaller than that of Keck, JWST offers a highly stable point spread function (PSF) and much greater sensitivity at wavelengths where cold BDs are brightest. For the JWST data reported here, the best combination of spatial resolution and SNR in the JWST data comes in the F360M filter. A simple Gaussian fit of the image of \wbd\ compared to fits of  two nearby stars shows no evidence for extent: FWHM(\wbd)= $0.150\pm0.001$\arcsec\ compared with $0.153\pm0.001$\arcsec and  $0.150\pm0.001$\arcsec for the two stars.

We explored the data further by constructing realistic models of the PSF and searching for a potential companion at all separations within 0.25".  \citet{Anderson2000} and \citet{Anderson2016} developed a novel technique applicable to multiple instruments on HST where they construct an ``effective'' PSF (ePSF) from bright single stars within a given image.  The ePSF is intended to describe the contemporaneous realization of the theoretical PSF on the detector pixels given the wavefront during the observation.  They build the ePSF first by identifying a sample of many bright stars in the image that appear to be singles and without contamination from cosmic rays or other stars.  These stars are all centered at various positions within their peak pixel, giving many samplings of the distribution of flux of the center of the PSF.  The initial ePSF is then generated by interpolating each PSF by the factor of user-defined oversampling, and taking the median.  Then, an iterative process begins, comparing the ePSF to all stars in the sample, evaluating the median residual, and adding that on to the ePSF model.  For our case, we generated an ePSF model 4x oversampled, as recommended in \citet{Anderson2016}, after 20 iterations of the ePSF calculation, over 9x9 detector pixels.  This analysis was performed using publicly available PSF building tools from the python package photutils \citep{Bradley2020}.

We chose the F360M data to perform this analysis as the number of bright sources in the F470N data was small which limits the production of a reasonable ePSF model.  Also, the diffraction limit in F470N is 30\% larger than in F360M, limiting the sensitivity to close companions.  We also avoid the F335M filter due to the low signal to noise of \wbd.  Of the four  pointings, two had reliable data for \wbd\ without nearby cosmic rays or bad pixels contaminating its flux.  Within each integration, we constructed a separate ePSF model from 57 and 63 separate stars (integration number 2 and 4, respectively), excluding \wbd\ from the sample.

\begin{figure*}[t!]%
\includegraphics[width=0.95\textwidth]{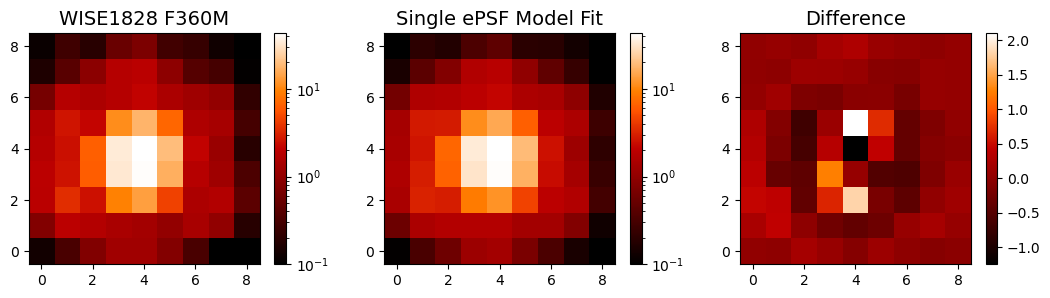}
\includegraphics[width=0.95\textwidth]{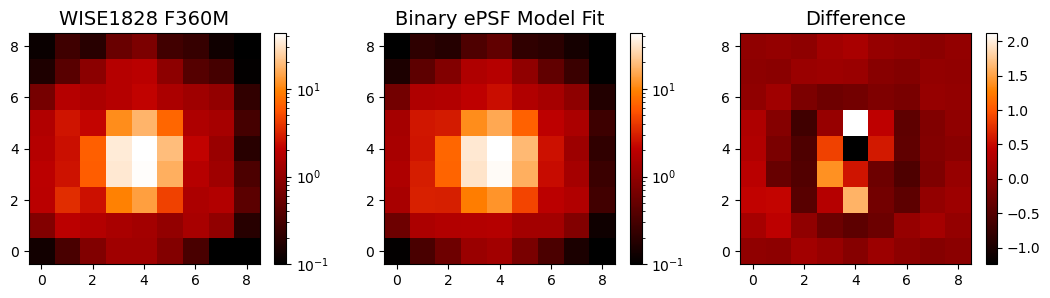}  
\caption{Top row: \wbd\ cutout in F360M filter on the left, ePSF single model in center, and residuals on the right. Bottom row: \wbd\ cutout in F360M filter on the left, ePSF binary model in center, and residuals on the right.  Units are in DN/s and the axes are in detector pixels (0.063"/pixel).  We used the level 2 pipeline product *cal.fits files to perform this analysis. \label{fig:wise1828binaryfit}}
\end{figure*}

\begin{figure*}[t!]
\includegraphics[width=0.85\textwidth]{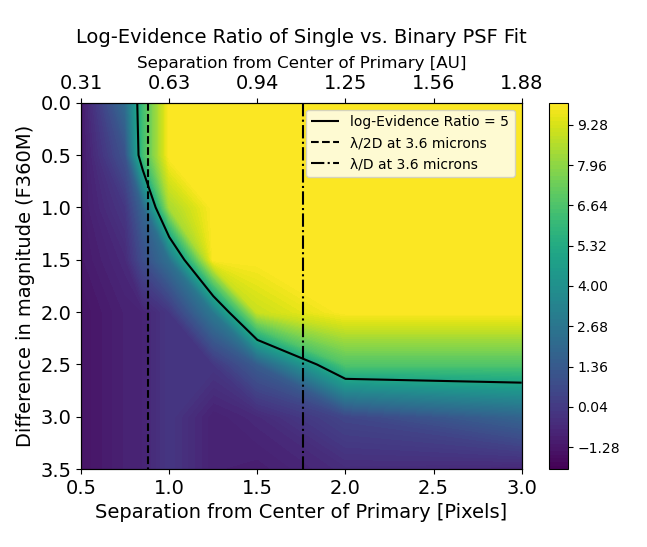} 
\caption{ Shown is the difference in log-evidence between the binary-PSF model and single-PSF model performed on artificial binaries, made from the ePSF models.  The threshold for strong evidence of the binary model fit over the single model fit is shown with a solid line (corresponding to a difference of 5).  The diffraction limit and half the diffraction limit at 3.6 microns are shown for comparison as the vertical dash-dotted and dashed lines, respectively. \label{fig:binary}}
\end{figure*}

We then made a double-PSF fitting code that takes a cutout array of the data as input (here a 9x9 pixel array centered around \wbd) and then fits the best-fit binary model using the ePSF, an approach similarly implemented on HST \citep{DeFurio2019}.  We use the python module PyMultiNest \citep{Buchner2014} that performs the Nested Sampling Monte Carlo analysis using MultiNest \citep{Feroz2009} to derive the best-fit binary ePSF model to the data, by maximizing our chi-sqaured likelihood statistic.  Our model consists of six parameters: x and y center of the primary, flux normalization of the primary, separation between the centers of the primary and secondary, the position angle of the center of the secondary relative to the primary, and the difference in magnitude between the secondary and primary.  We define flat priors with -1.5 $\leq$ x$_{cen}$ $\leq$ 1.5, -1.5 $\leq$ y$_{cen}$ $\leq$ 1.5, 0.0 $<$ flux normalization $<$ 40.0, 0.01 $\leq$ separation $\leq$ 4.0 pixels, 0.0 $\leq$ position angle $<$ 360.0, and 0.01 $\leq$ difference in magnitude $\leq$ 6.0 mag.

To both images of \wbd\ in F360M in question, the code converges to the edge of the prior in separation (4 pixels) and approaching that in difference in magnitude (5.6 and 5.8 mags in each).  If instead we force the code to fit within the core of the PSF (0.01 $\leq$ separation $\leq$ 2.0 pixels) and exclude wide separations, the best fit converges to 0.14$^{+0.04}_{-0.03}$ pixels in separation and $\Delta$mag=2.44$^{+0.34}_{-0.36}$  in Frame \#2, and 0.28$^{+0.13}_{-0.06}$ pixels in separation and $\Delta$mag=3.06$^{+0.74}_{-0.68}$ in Frame \#4 (errors are 68\% confidence interval). See Figure~\ref{fig:wise1828binaryfit} for the comparison of our binary PSF model to the data.

\begin{figure*}[t!]
    \includegraphics[width=0.5\textwidth]{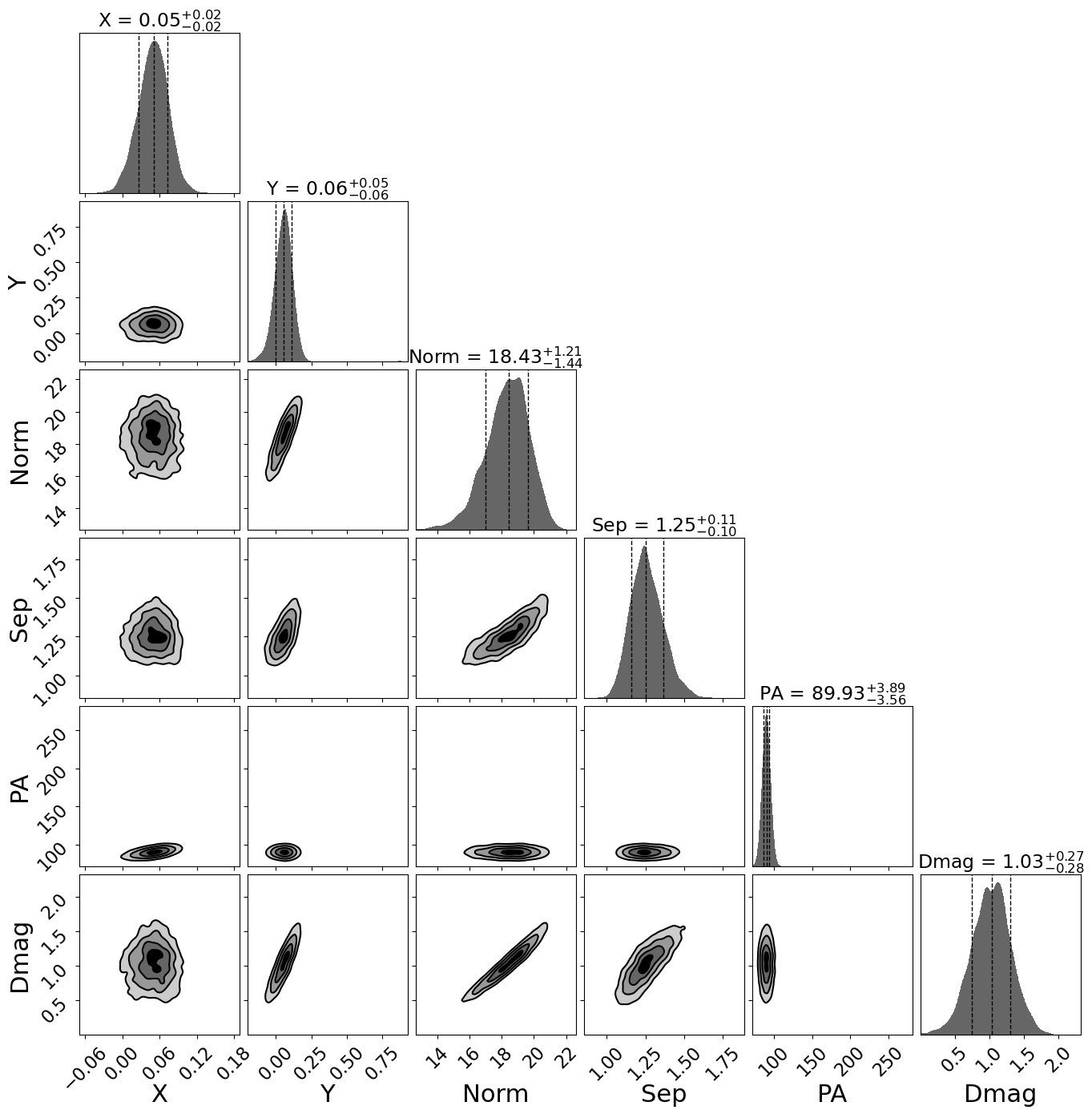} 
    \includegraphics[width=0.5\textwidth]{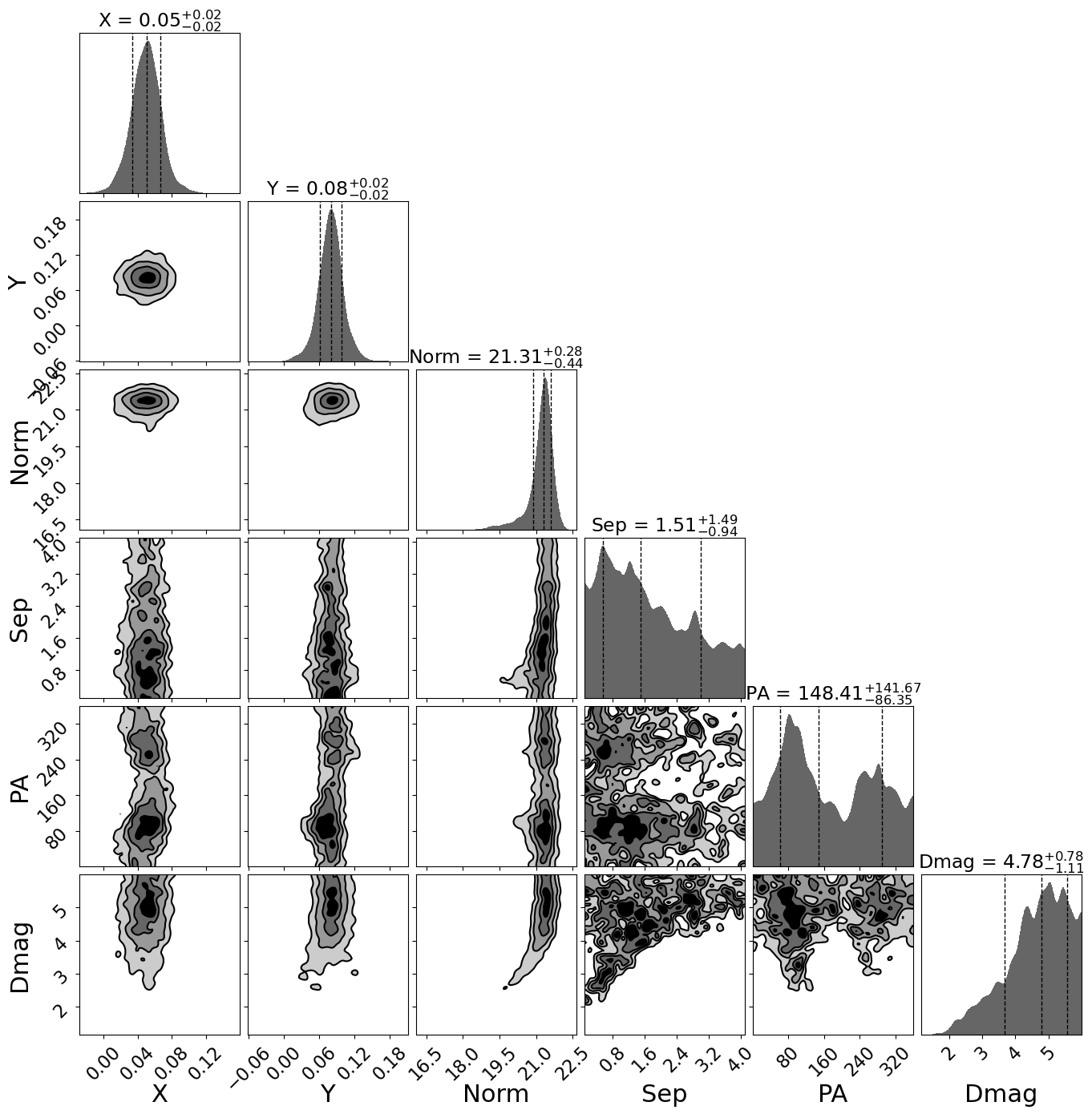} 
\caption{Binary model parameters from the recovery of a companion injected at 1.25 pixels in separation (Sep), a position angle (PA) of 90 degrees, and a contrast (Dmag) of 1.0 (left) and 3.5 (right) mag relative to the primary. These artificial binaries were constructed from empirical PSFs of NIRCam in the F360M filter.  X and Y are the central pixel coordinates of the primary and Norm is the flux normalization of the primary. \label{fig:cornerplots_binary}}
\end{figure*}

However, our binary PSF-fitting code will always find a best fit binary solution regardless of whether a true binary is present. If the object in question is a single point source, the code will either fit a companion to the brightest residuals in the background (e.g. the 4 pixels in separation and $\Delta$mag $\sim$ 5.5-6 mag initial fit) or to the residuals in the core of the source as models are photon-noise limited and have uncertainty (e.g. the resulting best fit when forced to fit a companion at separations $<$ 2 pixels).

In order to determine whether we can recover a companion with these values, we constructed many artificial binaries from the ePSF models at various separations and differences in magnitude with the same signal to noise as our F360M data.  Then, we ran our double-PSF fitting code and a modified version that just fits a single PSF (3 variables: x$_{cen}$, y$_{cen}$, and flux normalization) on those same artificial binaries.  PyMultiNest also calculates the evidence of the model in question by integrating over the posteriors.  \citet{Trotta2008} define a difference in the log-evidence between two models of 5 as being strong evidence for one model over another (with a probability of 0.993 that the higher evidence model is preferred).  For a given companion at some separation and difference in magnitude, we compare the log-evidence of the binary-PSF model to that of the single-PSF model. 

In Figure~\ref{fig:binary}, we show the log-evidence difference between the binary and single ePSF model fits based on separation and difference in magnitude of our artificial binaries.  We define our sensitivity as the point where the difference in log-evidence between the binary and single PSF models equals 5.  With our binary fitting tool and ePSF models, we can resolve companions down to $\sim$ 0.8 F360M  pixels (0.05\arcsec) in separation for an equal brightness companion and are sensitive to companions at $\Delta$mag=2.5 beyond 2 F360M pixels (0.126\arcsec).  See Figure~\ref{fig:cornerplots_binary} for the posteriors for a fit to a binary at 1.25 pixels in separation and $\Delta$mag=1.0 and 3.5 .  For the $\Delta$mag=3.5 companion, the parameters (separation, position angle, and difference in magnitude) are unconstrained. 

In addition, we fit a single PSF model to the \wbd\ data set. The difference in log-evidence between the single and binary PSF models is 3.5, less than the threshold for a detection. In Fig. \ref{fig:wise1828binaryfit}, we show both the single and binary PSF models compared to the data, demonstrating how the binary fit does not significantly improve the residuals. \textit{Therefore, we are confident that the best-fit binary model solution to \wbd\ in the F360M filter  is not a true detection of a companion, and that we can rule out an equal mass companion beyond 0.5 au and a $\Delta$mag=2.5  companion beyond 1.25 au.} Fainter companions at larger separations can also be ruled out although these would not help resolve the problem of excess brightness of \wbd\ relative to model predictions. Such models generally require an equal mass system as discussed below ($\S$\ref{sec:SED}).

The incidence of binarity among BDs is generally low,  10\%-30\%, compared with  higher mass stars \citep{Burgasser2007,Raghavan2010}.   While less is known about the multiplicity of the coldest  T/Y BDs, there is evidence that the  incidence of binarity may be even lower than for warmer BDs, with \cite{Opitz2016} failing to find any companions in the range of 0.5-2 AU among the five Y dwarfs they examined with AO imaging. Similarly, \citet{Fontanive2018}  estimated a companion frequency $<$ 10\% for T-Y dwarfs with separations tightly peaked at 3 AU, e.g. Luhman 16 AB \citep{Luhman2013}. However, it should be noted that separations $\leq$ 3 AU are not well explored.  If indeed \wbd\ is a binary as inferred from the spectral modeling presented by \citet{Cushing2021}, then high resolution JWST NIRSpec spectra may reveal a double lined system, given an appropriate orbital plane inclination. Furthermore,  two 5-10 \mj\ objects orbiting at 0.5 AU would have a period of 3-5 years and a typical orbital velocity of $\sim$5 km/s which might be discernible in multiple epochs of high SNR spectra to yield masses for the two objects.

\section{The Spectral Energy Distribution and Model Fits\label{sec:SED}}

The broad wavelength coverage of JWST NIRCam photometry from 1 to 5µm is ideal for characterizing the spectral energy distribution (SED) and inferring the bulk atmospheric properties of BDs.
We include thirteen ground- and space-based photometric points (i.e, Table \ref{tab:data}) in our spectral energy distribution fitting.
We considered a number of models for this comparison: Sonora Bobcat and  Cholla \citep{Marley2021}, ATMO \citep{Phillips2020} chemical-equilibrium (CEQ) models, ATMO chemical non-equilibrium models with strong vertical mixing (CNEQ-strong),  and the \citet{Linder2019} models.  
 We selected cloudless Sonora models\footnote{https://zenodo.org/record/1309035\#.YwqSGbTMKUk}$^,$\footnote{https://zenodo.org/record/5063476\#.YwqSTrTMKUk} as the most up to date compared, e.g. with older COND models, or the recent \citet{Linder2019} models which are limited to low mass objects ($<$2 \mj). The Sonora Cholla models do  not go to sufficiently low temperatures for this study and the ATMO 2020 models gave similar results but with higher reduced chi-squares than the Sonora Bobcat models.
 
 We defer discussion of the new \citet{Leggett2021} models to a companion paper on the NIRSpec  low and high resolution spectroscopy (Lew et al in preparation). \citet{Leggett2021} noted that their new model failed to fit the total luminosity of \wbd\ as a single object and assumed an equal mass binary.
 
We used the Sonora model grid that comprises temperatures from 200 to 600 K, gravities from log(g) of 3.0 to 5.5 $\, \mathrm ms^{-2}$, metallicities [M/H] from -0.5 to 0.5.
The ATMO model grid assumes an atmosphere with solar metallicity and span a range of temperature from 100\,K to 900\,K and of gravity from $10^{2.5}\, \mathrm cms^{-2}$ to $10^{5.5}\, \mathrm cms^{-2}$.
The ATMO chemical non-equilibrium models with strong vertical mixing assume an eddy diffusion coefficient of $10^{6} \,\mathrm cm^2s^{-1}$ at a gravity of $10^{4.5}\,\mathrm cms^{-2}$ and decreases with higher gravity (see Figure 1 in \citealt{Phillips2020}.
We linearly interpolated the models to construct a model spectra grid with a smaller spacing in temperature, gravity, and metallicity.
We adopt a distance of 9.93\,pc reported by \citet{Kirkpatrick2019}.

In the model fitting process, we examined two cases: a single object or a binary with different temperatures.
For the SED fitting, we performed the least-squares fit to the thirteen photometric points.
For the binary case, we require the two objects to share the same age and metallicity. 
To enforce the equal-age constraints, we first calculate the age based on the free parameters temperature $T_1$ and gravity $g_1$ of one component of the binary using the Sonora evolution model.
We then calculate the expected gravity $g_2$ of the second component based on the calculated age and sampled temperature $T_2$ of this object.
Therefore, both components of the binary share the same age, where $g_2$ is thus a derived value based on $T_1,g_1,T_2$ and $M$.
We calculate the chi-squared values of the binary models over the grid of temperature, gravity, metallicity, and radius of the binary components.
We use the bootstrapping method to estimate the uncertainties of the fitted parameters. 
With the bootstrapping method, we randomly re-sample the thirteen photometric points for 10,000 times and refit the models to the resampled datapoints with the least squares method.
We then calculate the 99.7 percentile range of the fitted parameters as the 3$\sigma$ confidence ranges, as shown in Table \ref{tab:Sonora}.

\begin{figure*}
    \centering
     \begin{minipage}{0.35\textwidth}
 \includegraphics[width=1.1\textwidth]{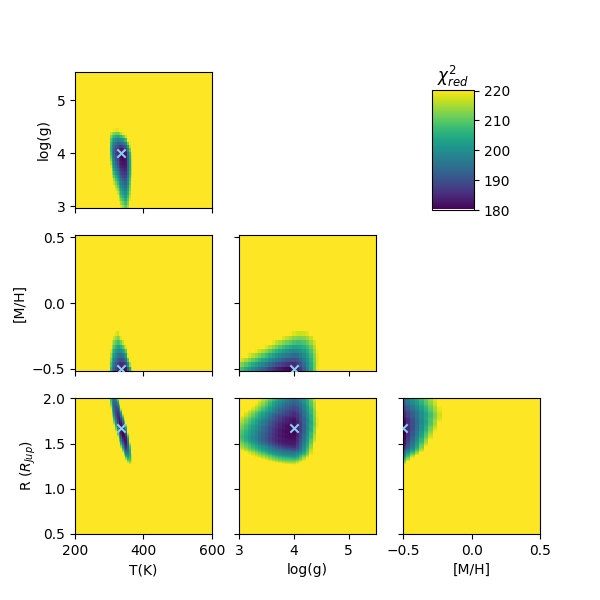}
 \end{minipage}
 \begin{minipage}{0.64\textwidth}
 \includegraphics[width=1\textwidth]{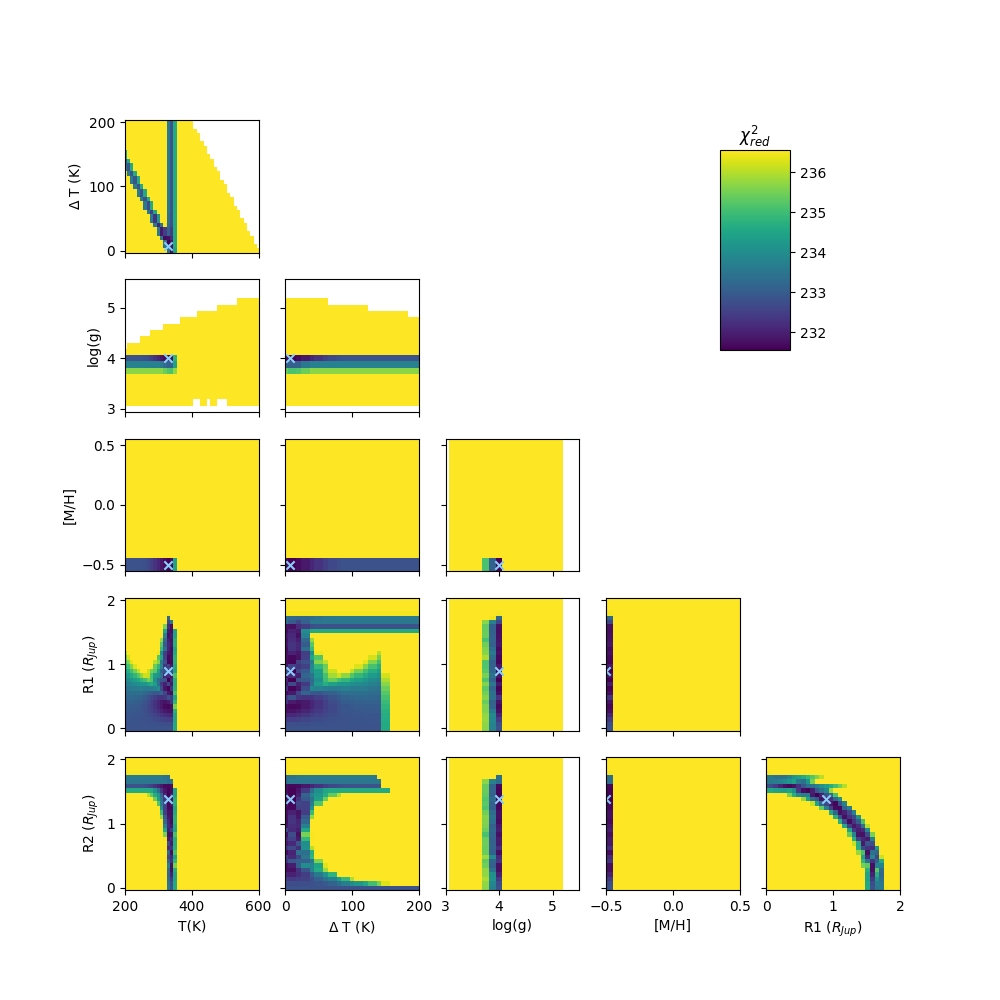}
 \end{minipage}
    \caption{The reduced chi-squared maps of the photometry fitting results for both the single object case (left) and the binary case (right). Both models fitting results suggest \wbd\ has a temperature of around 330K, low metallicity, and low gravity. The blue crosses mark the location with the lowest chi-squared values that are listed in Table \ref{tab:Sonora}. The white-colored regions in the plots are either outside of the considered temperature range (i.e., T + $\Delta T>$600K) or incompatible with the Bobcat evolution model grid.}
    \label{fig:MCMC}
\end{figure*}

In Table \ref{tab:Sonora} and Figure \ref{fig:MCMC}, we show the best-fit parameters and display the corner plots associated with the fits to show the correlation between the parameters in our SED fitting.
In the binary case, the chi-squared map suggests that the primary shares a similar temperature ($\Delta T<$50K) to the secondary when both components have non-negligible radii ($R_1$ and $R_2 > 0$).
We also calculate the combined radius, which is the square root of the sum of the two radii squared ($\sqrt{R_1^2+R_2^2}$), for the binary case. 
The combined radius is moderately constrained, with a 99.7 percentile range of 1.15-2.83 $R_J$ and a median value of 1.67$R_J$, where the best-fit radius of the single model is 1.67$R_J$.
Both the single and binary fits favor effective temperature of around 330K, low metallicity ([M/H = -0.5]), and low gravity ($\log(g) = 4$).
However, the best fit models suggest that the single model requires a lower mass object (2\mj), while the binary model requires two higher mass objects ($\sim$4\mj).  Our single model has a lower reduced $\chi^2$ than the binary model, although both are too high to adequtely represent the data. We note that the best-fitted radius of 1.67 $R_J$ in the single BD case may be unphysically large, and that an equal-temperature binary may be preferred in this case, as suggested in previous studies of \wbd\ \citep[e.g.,][]{Beichman2013,Leggett2013,Leggett2017,Cushing2021,Leggett2021}


A striking characteristic of the SED fits for both the single and binary cases is their predicted youth, with best-fit ages between 0.3-0.7 Gyr which is at odds with the likely dynamical age of \wbd, although these young ages are not well constrained in our model. As previously noted   \citep{Beichman2014}, \wbd\ is not associated with any known young cluster. This original conclusion is bolstered by application of the Banyan $\Sigma$ tool\footnote{www.exoplanetes.umontreal.ca/banyan/banyansigma.php?} \citep{Gagne2018} which puts the probability that \wbd\ is a field object at 99\% independent of its as yet unknown radial velocity. We expect \wbd\ to have an age in the 2-4 Gyr range given its tangential velocity \citep{Beichman2013}, consistent with the mean age of 2.3 Gyr for a sample of BDs in the solar neighborhood \citep{DupuyLiu2017}, contrary to our modeling of the SED. 

Based on the best-fit models, we extrapolate the flux density beyond the observed wavelength regions and estimate the bolometric luminosity. The estimated bolometric luminosity has $\log(L/L_{\odot})$ of around -6.5, as listed in Table \ref{tab:data}. The bolometric luminosity of $\log(L/L_{\odot}) =$ -6.5 is similar to the expected luminosity of a $\sim$400K object at 1Gyr. Our calculations indicate that the JWST broadband photometry covers about 16\% of the bolometric luminosity, while the composite photometry comprising JWST, 2MASS, WISE, and Spitzer photometry accounts for around 55\% of the bolometric luminosity.



The discrepancy between models for cold BDs at long and short wavelengths is well known \citep{Beichman2014} with  models which fit well at long wavelengths failing to fit at short wavelengths and vice versa. This  failure is prominent in our fits to the \wbd\ data, independent of metallicity or multiplicity. In all cases,  the models, which fit reasonably at 4-5 \mum, fail badly at 1 \mum\ with the models being a factor of $\sim$2 brighter than observed at 1 \mum. Another striking feature is that both the F335M and F360M data points sit significantly above the predictions suggesting that the absorption in this part of the spectrum is less than expected. Further exploration of the possible temperature-pressure profile and atmospheric chemistry, such as those models in \citet{Leggett2021}, is essential to understand the atmospheric processes that drive the longstanding challenges in fitting the flux at near-IR and mid-IR wavelengths.



The SED fits to new JWST observations using two independent sets of models leave us in the uncomfortable position of either accepting a single object much younger than expected dynamically, or a near equal-mass binary system for which the NIRCam data finds no evidence at separations $>$0.5 AU.  Potentially the NIRSpec's high resolution spectroscopy will reveal a double-lined system which might resolve this problem. However, whether \wbd\ turns out to be  a double,  Figure~\ref{fig:photometry} and the high $\chi^2$ values indicate significant deviations between  the models and the emission of \wbd. 

At the fitted effective temperature of around 300K, it is possible that sulfide, chloride, and water clouds form near or within  the photosphere and affect the emission spectra \citep[e.g.,][]{Morley2012,Morley2014a, Morley2014b}.
However, \cite{Cushing2021} SED fitting results with \citet{Morley2014a} cloud models at solar metallicity give a higher reduced chi square than that with Sonora bobcat cloudless models with subsolar metallicity. 
Further exploring cloudy models under various atmospheric chemistry and metallicity is essential to understand the role of clouds in shaping the SED of  WISE 1828+2650. It is possible that the rotation rate and observed inclination of \wbd\ cause significant luminosity deviations from a typical BD of similar temperature. Deviations as high as 20\% have been estimated in comparing pole-on vs. equator-on viewed BDs at high rotation rates \citep{Lipatov2022}, but it is currently unknown if this effect occurs for \wbd\ . Important clues will come from  the NIRSpec and MIRI  spectroscopic observations spanning 1-12 \mum\ at low and high resolution which are currently being analyzed. 
 


\begin{deluxetable*}{l|llll|llll|lllll}[ht]
\tabletypesize{\scriptsize}
\tablewidth{0pt}
\tablecaption{Spectral Model Fits to Photometry\label{tab:Sonora} }
\tablehead{
\multicolumn{2}{c}{} &
\multicolumn{3}{c}{Primary} &
\multicolumn{3}{c}{Secondary}\\
\colhead{} & \colhead{T$_{eff}$} & \colhead{log g} & \colhead{Mass} & \colhead{Radius}  & \colhead{T$_{eff}$} & \colhead{log g} & \colhead{Mass} & \colhead{Radius} & \colhead{Metallicity} & \colhead{Age} & \colhead{Reduced} & \colhead{Degs. of} & \colhead{$\log$ ($L/L_{\odot}$)}\\
\colhead{}&\colhead{(K)}&\colhead{(cgs)}& \colhead{(M$_{Jup}$)} & \colhead{(R$_{Jup}$ )} &\colhead{(K)}&\colhead{(cgs)}& \colhead{(M$_{Jup}$)}& \colhead{(R$_{Jup}$)} & [M/H] &\colhead{($\times10^{9}$yr)} & $\chi^{2}$& \colhead{Freedom} &\colhead{}}
\startdata
\multicolumn{14}{c}{Sonora Cloudless models} \\
\hline
best-fit single & 325 &3.6 &2 &1.83 &&&&&-0.5$^*$&0.3 &180&9 & -6.49$\pm$ 0.03\\
99.7$\%$ CI & 309--405 & 3.0--5.5&0.6--13 &1.1--2.0&&&&&-0.5 &0.01--7&&& \\
\hline
 best-fit binary & 337& 4.0 & 4.4 &1.38 & 330 & 4.0 & 4.1 & 0.90 &-0.5$^*$&0.7& 231 &7  &-6.49$\pm$0.04 \\
99.7$\%$ CI & 316--413 & 3.0--4.8 &0.6--20&0.4--2& 268--395 & 3.5--4.8 & 0.6--20 & 0--2 &-0.5& 0.03--10&&&\\
\hline
\multicolumn{14}{c}{ATMO Chemical Equilibrium models} \\
\hline
best-fit single &300 &4.0 &3.7 &2 &&&&& 0 (fixed) &1.0 &256 &10 & -6.45$\pm$ 0.06\\
99.7$\%$ CI & 287--359 &3.0--5.1 &0.5--13 &2--2 &&&&&- &0.02--8.5 &&&\\
\hline
 best-fit binary & 300 & 3.6 & 1.5 &1.9 & 300 & 3.6 & 1.7 & 1.4 & 0 (fixed) &0.3 & 287 &8  &-6.36$\pm$0.05 \\
99.7$\%$ CI & 279--400 & 3.0--4.8 &0.5--20 &0.4--2 &279--400 &3.0--4.8 & 0.5--20 & 0--2 & - & 0.02--10&&&\\
\hline
\multicolumn{14}{c}{ATMO Chemical Non-equilibrium models with strong mixing} \\
\hline
best-fit single & 294 &4.3 &5.5 &2.0 &&&&& 0 (fixed) &2.4 &303 &10 & -6.46$\pm$ 0.06\\
99.7$\%$ CI & 259--359 & 3.0--5.0 &0.5--11 &2.0--2.0&&&&&- &0.02--9.9 &&&\\
\hline
 best-fit binary & 270& 4.0 & 4.2 & 2 & 270 & 4.0 & 4.2 & 2 & 0 (fixed) &2.1 & 320 &8  &-6.38$\pm$0.09 \\
99.7$\%$ CI & 253--340 & 3.0--4.6 &0.5--13 &2--2& 253--340 & 3.0--4.6 & 0.5--13 & 2--2 &-& 0.03--9.9&&&\\
\enddata
\tablecomments{$^1$ The fitted parameters and the corresponding 99.7$\%$ percentile ranges for the single and binary objects cases. The mass, radius, and the secondary object's gravity are derived from the fitted parameters using the Bobcat evolution models. $^*$The fitted metallicity is at the lower bound of metallicity grid.}
\end{deluxetable*}



\begin{figure*}[t!]%
\includegraphics[width=0.835\textwidth]{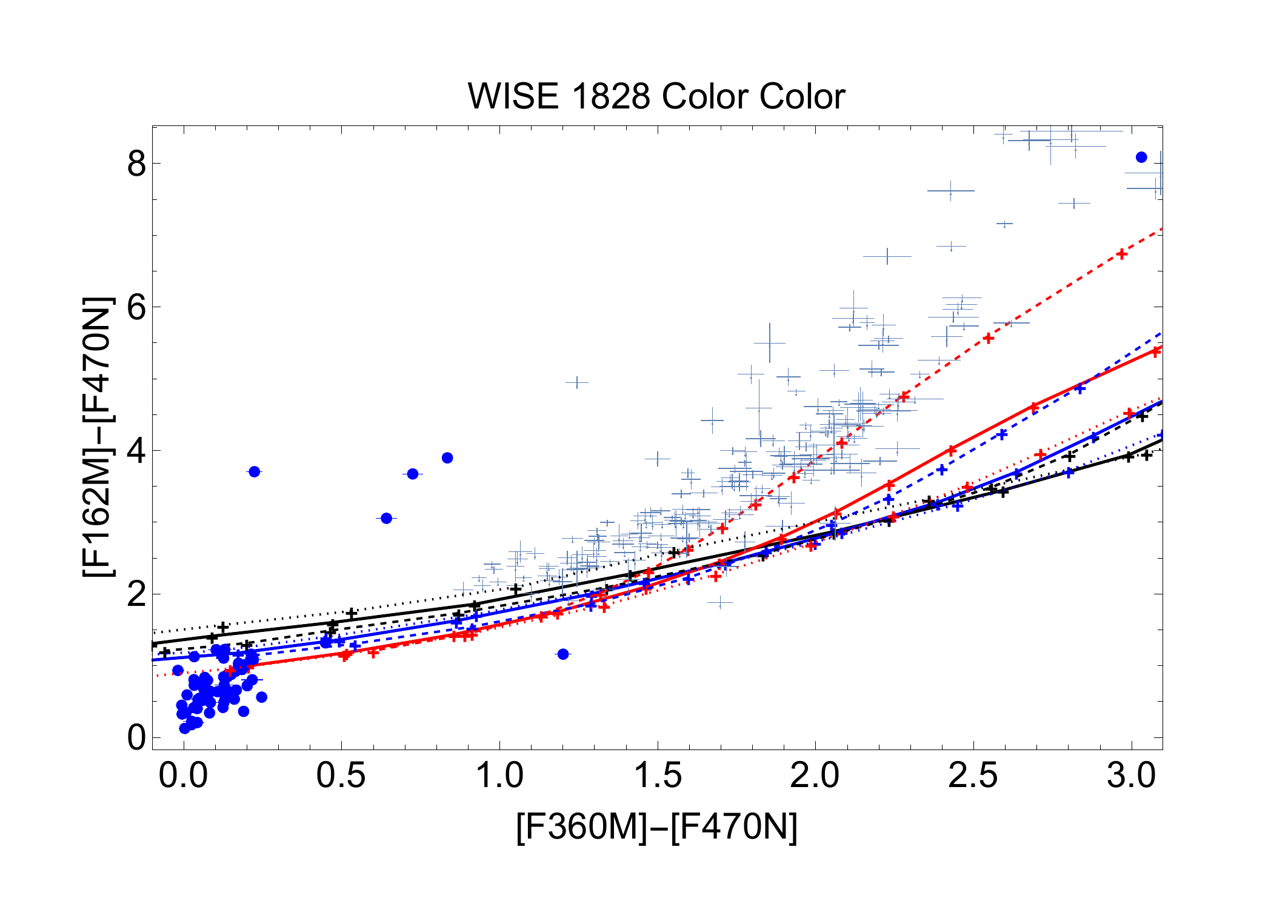} 
\caption{ Color-Color plot for point sources  detected in all three bands seeded with a high SNR detection in F360M. Blue circular symbols are sources in the \wbd\  field (\wbd\ is in the upper right corner), the light blue crosses represent the progression of  BDs from \citet{Kirkpatrick2019} plotted in using H band, Spitzer IRAC Ch1 and Ch2  similar  to the JWST filters.  Visual examination of the F360M image shows that four JWST sources that lie above the BD locus are  slightly extended, very red galaxies. The lines denote sequence of early L to late T dwarfs based on Sonora Models \citep{Marley2021}. The colors denote values of log(g) (black, 3.0, blue, 4.0, red, 5.0). The line type denotes different values of [Fe/H]=(0,thick; -0.5,dashed; and +0.5 dotted).    \label{fig:ColorColor}}
\end{figure*}

\section{A Search for Brown Dwarf Companions in the Entire Field \label{sec:widefield}}

The large field of view and great sensitivity of NIRCam are two of the instrument's great strengths. In addition to searching for a close companion we also looked for additional BD candidates which might or might not be associated with \wbd. For example,  \citet{Nonino2022} made a serendipitous discovery of a  T dwarf (T$_{eff}\sim$600 K) at a distance of $\sim0.5$ kpc in a deep survey field adjacent to  Abell 2744.

We used the source catalogs provided by Level-3 processing from STScI for  our 6 filters. We selected F360M unconfused point sources (the "is\_extended" flag set to FALSE and the nearest neighbor distance $\geq$ 1\arcsec) with SNR$\geq$10. To avoid unreliable sources which appear  at the edges of the field due to having less than the full coverage of 4 dither positions, we selected objects only within  the central $2^\prime\times2^\prime$ of the final F360M image.  We selected sources  in the other 5 filters (SNR$\geq$5) in the same way and used the F360M  objects as seeds for  band merging using a radius of 0.2\arcsec. We adopted the catalog values of aperture photometry based on 70\% encircled value with pipeline-provided aperture correction. To enable a search for L,T and Y BDs we focused on filters optimized to find  cold objects, requiring detections at F360M, F162M and F470N. Figure~\ref{fig:ColorColor} shows  63 sources meeting these criteria. Sources around 0.5$>$[F360M]--[F470N]$>$0  and $0<$ [F162M]--[F470N] $<2$ mag are likely stars with spectral types as cool as M9 or galaxies \citep{Pecaut2013}\footnote{\url{https://www.pas.rochester.edu/~emamajek/EEM_dwarf_UBVIJHK_colors_Teff.txt}}.

Figure~\ref{fig:ColorColor} shows  only one  object in the color space occupied by BDs,  \wbd\ itself in the upper right corner of the plot.  Finally, there are 4 other  red objects with colors around  [F162M]--[F470N] $>$2 mag (Table~\ref{tab:3band}). A sixth source with [F360M]--[F470N]$\sim 1.2$ lies below the BD locus and appears to be  contaminated by confusion with a nearby object.  Visual inspection and measured FWHM of the four red objects shows two to be   extended (Table~\ref{tab:3band}). A more detailed examination of the images and comparison with the predicted WebbPSF\footnote{\url{https://webbpsf.readthedocs.io/en/latest/}} image size shows that  \#2 and \#4 have  FHWM  in the F360M image  of 0.23\arcsec and 0.43\arcsec, respectively, compared with other point sources in the field (0.15\arcsec in the mosaicked images) and the WebbPSF FHWM of 0.12-0.13\arcsec (Table~\ref{tab:3band}). A third, \#5, may be slightly extended 0.18\arcsec. The other two, \#1 and \#6, are point-like or only very slightly extended. Source \#3 is \wbd. Extra-galactic objects with similar colors are being found in on-going deep imaging programs and are suggestive of a new class of highly dust obscured galaxy (Hainline et al, in preparation).

Figure~\ref{fig:ColorColor}  includes colors from a large sample of BDs from \citet{Kirkpatrick2021} using H-band, IRAC C1 and Ch2 as analogs for the JWST filters. The plot also includes  loci of six different Sonora BD models evaluated in the JWST filters and ranging in surface temperature, gravity, and metallicity: $300<T_{Eff}<1700$ K, log(g) of (3,4,5), and [Fe/H] of ($-$0.5,0,+0.5). The comparison between the models and the Kirkpatrick sample  show the  well known problem that as the BDs become cooler, the models have a progressively harder time fitting the data. 

The conclusion of this search and visual inspection of the six JWST images is that there are no obvious cool BD candidates in the \wbd\ field, but that there are a significant number of galaxies.

\begin{deluxetable*}{r|ll|llllll|l}
\tablecaption{Highly Red Sources in \wbd\ Field \label{tab:3band} }
\tablehead{
\colhead{Label}&\colhead{RA2000$^1$}&\colhead{DEC2000$^1$}&\colhead{F090W$^2$}  &\colhead{F115W$^2$}&\colhead{F162M$^2$}&\colhead{F335M$^2$} &\colhead{F360M$^2$} &\colhead{F470N$^2$}&\colhead{FWHM$^3$}}
\startdata
1 &277.138408(60.) &26.8286435(55.)& 1.06$\pm$0.00769 & 4.08$\pm$0.00955& 0.911$\pm$0.00908&  6.23$\pm$0.0223 &5.66$\pm$0.0218& 4.24$\pm$0.0982&0.15\arcsec\\ 
\, \,&18h28m33.22s&26d49m43.12s& 23.33$\pm$0.0078&  21.59$\pm$0.0025 &22.62$\pm$0.011 & 19.2$\pm$0.0039 &19.15$\pm$0.0042& 18.92$\pm$0.025& \\ 
2 &277.160802(20.) &26.8420123(16.)& 0.172$\pm$0.00651 & 0.545$\pm$0.00633& 0.975$\pm$0.00975&  2.$\pm$0.0113 &2.27$\pm$0.0122& 2.5$\pm$0.0747&0.23\arcsec\\ 
\, \,&18h28m38.59s&26d50m31.24s& 25.3$\pm$0.04&  23.78$\pm$0.013 &22.55$\pm$0.011 & 20.43$\pm$0.0061 &20.14$\pm$0.0058& 19.5$\pm$0.032&\\ 
3$^4$ &277.133408(26.) &26.8444326(4.3)& 0.078$\pm$0.005 & 0.311$\pm$0.00446& 1.29$\pm$0.00857&  2.41$\pm$0.0132 &34.2$\pm$0.0511& 340.$\pm$0.619&0.15\arcsec\\ 
\, \,&18h28m32.02s&26d50m39.96s& 26.16$\pm$0.20 &  24.39$\pm$0.015 &22.25$\pm$0.0072 & 20.23$\pm$0.0059 &17.19$\pm$0.0016& 14.16$\pm$0.002&\\ 
4 &277.16172(24.) &26.8470549(26.)& 0.0796$\pm$0.00708 & 0.323$\pm$0.00663& 0.562$\pm$0.00985&  1.93$\pm$0.0112 &2.14$\pm$0.012& 2.55$\pm$0.0752&0.44\arcsec\\ 
\, \,&18h28m38.81s&26d50m49.4s& 26.14$\pm$0.093&  24.34$\pm$0.022 &23.15$\pm$0.019 & 20.47$\pm$0.0063 &20.2$\pm$0.0061& 19.48$\pm$0.032&\\ 
5 &277.138946(24.) &26.8479253(5.9)& N/A  & 0.23$\pm$0.0043& 1.27$\pm$0.00946&  4.85$\pm$0.0166 &5.39$\pm$0.0176& 7.1$\pm$0.098&0.18\arcsec\\ 
\, \,&18h28m33.35s&26d50m52.53s& N/A &  24.71$\pm$0.02 &22.26$\pm$0.008 & 19.47$\pm$0.0037 &19.2$\pm$0.0035& 18.36$\pm$0.015&\\ 
6$^5$ &277.152549(430.) &26.8653645(380.)& 2.42$\pm$0.00887 & 3.94$\pm$0.0116& 9.63$\pm$0.0276&  6.41$\pm$0.0218 &2.34$\pm$0.0225& 4.31$\pm$0.0968&0.15\arcsec\\ 
\, \,&18h28m36.61s&26d51m55.31s& 22.44$\pm$0.004&  21.63$\pm$0.0032 &20.06$\pm$0.0031 & 19.17$\pm$0.0037 &20.11$\pm$0.01& 18.91$\pm$0.024&\\ 
\enddata
\tablecomments{$^1$Epoch 2022.5699, uncertainties in milliarcseconds in parentheses; $^2$Top line is flux density in $\mu$Jy, the second line is corresponding Vega magnitude.$^3$Full width at Half Maximum in F360M. This is to be compared with the FWHM of the WebbPSF of 0.17\arcsec. $^4$Source is WISE1828+2650. $^5$Source to confusion with a nearby object.}
\end{deluxetable*}


\section{Conclusion\label{sec:conclusion}}

This examination of the new 1-5\mum\ NIRCam data for \wbd\ has confirmed  what has been previously known about this source.  It remains  among the reddest and most challenging of the Y dwarfs to fit using existing models. Both the high SNR data at F360M and the data from the Keck telescope at 1.6 \mum\ have failed to reveal a near-equal mass companion beyond 0.5 AU. If, as has been suggested, a binary system offers a better chance of fitting the photometric models, the companion must orbit very close to \wbd\ itself, $<$0.5 AU. Such a companion might reveal itself through double lines in the spectroscopy. However, as described above, even the binary model fails to provide an improved fit to the existing photometric data. Further insights into the properties of this enigmatic object will come with the analysis of the NIRCam and MIRI spectroscopic data forthcoming shortly in other publications.

\section{Acknowledgements}
We must first acknowledge the many years of effort by thousands of scientists, engineers and administrators who made  JWST  such a dramatic success, exceeding many of its most important requirements. This international collaboration should be an inspiration to us all. The NIRCam team at the University of Arizona and Lockheed Martin's Advanced Technology Center brought great skill,  expertise and dedication to realize the full power of this instrument. M. De Furio is grateful for support from NASA through the JWST NIRCam project though contract number NAS5-02105 (M. Rieke, University of Arizona, PI).

We wish to thank Davy Kirkpatrick for valuable discussions about WISE 1828. Some of the research described in this publication was carried out at the Jet Propulsion Laboratory, California Institute of Technology, under a contract with the National Aeronautics and Space Administration. M. De Furio benefited from support from JPL's Strategic University Research Partnership (SURP). Doug Johnstone is supported by NRC Canada and by an NSERC Discovery Grant. L.A. acknowledges support by the Canadian Space Agency under contract 9F052-170914/001/MTB.

Some of the data presented herein were obtained at the W. M. Keck Observatory, which is operated as a scientific partnership among the California Institute of Technology, the University of California and the National Aeronautics and Space Administration. The Observatory was made possible by the generous financial support of the W. M. Keck Foundation.  The authors wish to recognize and acknowledge the very significant cultural role and reverence that the summit of Maunakea has always had within the indigenous Hawaiian community.  We are most fortunate to have the opportunity to conduct observations from this mountain.

\bibliography{main}{}
\bibliographystyle{aasjournal}

\end{document}